\documentstyle[12pt,a41,epsfig,wrapfig]{article}

\begin{document}
\begin{flushright}
DESY 97-224\\
hep-ph/9711290
\end{flushright}
\vspace*{1cm}

\begin{center}
{\Large \bf Double Spin Asymmetries in Charmonium  Hadroproduction 
at HERA-$\vec N$}\footnote{Talk given by A.~Tkabladze at
 the Topical Workshop 'Deep
Inelastic Scattering off Polarized Targets: Theory Meets Experiment',
DESY-Zeuthen, September 1-5, 1997.} \\
\vspace*{0.5cm}
{W.-D.~NOWAK$^a$, O.~TERYAEV$^b$, A. TKABLADZE$^{a,b}$}\\
\vspace*{5mm}
{\it $^a$DESY-IfH Zeuthen, Germany}\\
\vspace*{3mm}
{\it $^b$Bogoliubov Laboratory of Theoretical Physics,}\\
{\it JINR, Dubna, Russia}

\vspace*{1cm}

\end{center}

\begin{abstract}
{\small
We discuss the   double spin asymmetries in charmonium hadroproduction
 with nonzero transverse momenta at
fixed target energies, $\sqrt{s}\simeq40$ GeV, within the framework of the
factorization approach.
It is shown that the color octet contribution is dominant in the asymmetries.
The size of the  asymmetries and the projected statistical errors
in a future option of HERA with longitudinally
polarized protons scattering off a polarized target (HERA-$\vec N$)
 should allow  to
distinguish between different parametrizations for the polarized gluon
 distribution in the proton.}
\end{abstract}

\section{Introduction}
\setcounter{equation}{0}

The study of spin asymmetries in the production of 
heavy quarkonium states  in
polarized nucleon-nucleon collisions should provide  important
information about the proton spin structure. Heavy quark-antiquark
production processes occur at  small distances and the subprocess level
cross sections can be calculated perturbatively.    
On one hand, charmonium production asymmetries are expected to be
sensitive to the polarized gluon distribution function in the proton,
since  heavy quark systems  are mainly produced in gluon-gluon fusion
subprocesses.
On the other hand, it is essential  to  investigate in more
detail  the heavy quark-antiquark pair hadronization phase. To this
end,  observation of charmonium production in polarized
experiments is expected to  provide  additional tests for existing models.

The double spin asymmetry in $J/\psi$ production has been studied in the
framework of the so-called color singlet model (CSM) \cite{CSM} by Morii and
collaborators \cite{MTY}. 
The CSM is a nonrelativistic model where the relative velocity, $v$, between
the heavy constituents in a bound state is  neglected. However, discrepancies
between experimental data 
[3-10] and the CSM predictions 
indicate  that $O(v)$
corrections as
well as other mechanisms  of quarkonium production, which do not appear
in the leading order in $v$, should be considered.
Such an  expansion of the quarkonium cross sections and decay widths
has been realized in the last few years within the framework of the 
Factorization Approach (FA)   based on  Nonrelativistic QCD (NRQCD) 
\cite{NRQCD}. Here 
the production cross section for a quarkonium state  H in the process
\begin{eqnarray}
 A+B\to H+X
\end{eqnarray}
can be written as
\begin{eqnarray}
  \sigma_{ij} = \sum_{i,j}{\int_{0}^{1} {dx_1 dx_2 f_{i/A}(x_1) f_{j/B}(x_2)
\hat\sigma(ij\to H)}},
\end{eqnarray}
where   $f_{i/A}$ is the distribution function of the
parton $i$ in the hadron $A$, and 
\begin{eqnarray}
\hat\sigma(ij\to H) = \sum_{n}{C^{ij}[n]\langle0|{\cal O}^{H}[n]|0\rangle},\nonumber
\end{eqnarray}
the subprocess  cross section, is separated into two parts: short
distance~   coefficients, $C^{ij}[n]$,
and~ long~ distance matrix elements, $\langle0|{\cal O}^H[n]|0\rangle$.
 The $C^{ij}[n]$ is the production
cross section of a heavy quark-antiquark pair by fusion of partons $i$ and $j$.
It can be calculated in the framework of pQCD. The [n] state can be
either  a color
singlet or a color octet state. 
The $\langle0|{\cal O}^H[n]|0\rangle$ describes the evolution
of a quark-antiquark pair into a hadronic state. These matrix elements cannot
be computed perturbatively, but the relative importance of long distance matrix
elements in powers of velocity $v$ can be estimated  using
the NRQCD velocity scaling rules \cite{LMNMH}.
 Unlike the color singlet long distance matrix elements, each connected
 with the subsequent
hadronic nonrelativistic wave function at the origin, color octet long
distance matrix elements are unknown and should be extracted from
experimental data.

The new  formalism implies that quark-antiquark color octet intermediate
 states are allowed
to contribute to heavy quarkonium  production and decay
processes at higher order in the velocity expansion.
Therefore, in the Color Octet Mechanism (COM)   the complete structure of
the quarkonium Fock space is taken into account while in the CSM
only the dominant Fock state is considered, which consists of a color singlet
quark-antiquark pair in a definite angular-momentum state of a final
hadron (the leading term in the velocity expansion).

The shape of the  $p_T$ distribution of the $^3S_1^{(8)}$ octet state
production cross section indicates that $J/\psi$ and $\psi'$
production at large $p_T$ observed at the Tevatron (FNAL)
 can be explained in the FA \cite{BF,CL}.  
The color octet contribution to  $J/\psi$ photoproduction has been analyzed
in  papers \cite{CK,AFM}. Recently,  $J/\psi$ hadroproduction at fixed
target energies has been studied by including the color octet mechanism
\cite{GuS,BeR,TV}. Large discrepancies between experimental data and the CSM
predictions for the total cross section of
$J/\psi$ hadroproduction were explained.
The color octet contribution is dominant in  $J/\psi$ hadroproduction at
energies $\sqrt{s}\simeq30\div60$ GeV.
The COM prediction for the ratio
$\sigma(J/\psi)_{dir}/\sigma(J/\psi)\simeq0.6$ is also in a good agreement with
experimental data \cite{Beneke}.

Despite the obvious successes of the COM  some problems 
remain unsolved.
 In particular, the theoretical
predictions disagree  with  the $J/\psi$ and $\psi'$ polarization data at
fixed target energies \cite{BeR,Beneke} and the COM prediction for the
yield ratio of $\chi_{c1}$ and $\chi_{c2}$ states remains too low \cite{BeR}.
These discrepancies
indicate that higher twist corrections may give a significant contribution
to low $p_T$ production of charmonium states and should be added
to the color octet contributions \cite{VHBT}.
The color octet contribution underestimates the $J/\psi$
photoproduction cross section at large values of $z$
($z=E_{J/\psi}/E_{\gamma}$ in the laboratory frame) \cite{Kramer}.

The NRQCD factorization approach implies universality, i.e. the values
of long distance matrix elements extracted from the different
experimental data must be the same. 
However, due to the presently rather large theoretical uncertainties, the
existing  experimental data 
does not allow  to check the FA universality and therefore to test the COM.

This fact motivated us to look for other processes with less
theoretical uncertainties to test the color octet mechanism.
The observation of $J/\psi$ asymmetries can be used for these purposes
as well as  measurements of  the $J/\psi$ polarization
in  unpolarized hadron-hadron collisions and electroproduction 
\cite{VHBT,BK, FleN}.
 
 In this talk we consider the possibility of extracting
information about the gluon polarization in the nucleon through spin
asymmetries in charmonium production. And conversely, 
we investigate what can be learned from charmonium production
asymmetries  about  color octet long
distance matrix elements, if the polarized gluon distribution
function should once be  measured in other experiments.

Finally, we will present the results of our calculation of the
expected spin asymmetries in production 
of $J/\psi$ and $\chi_{cJ}$ states 
 at HERA-$\vec N$, one of the future options of HERA \cite{Nowak};
an experiment utilizing an internal polarized nucleon target in the  polarized
HERA beam with energy $820$ GeV would yield $\sqrt{s}\simeq40$ GeV.
To avoid the uncertainties coming from higher twist subprocesses \cite{VHBT}
we considered $J/\psi$ production at large enough $p_T$ which can not be
caused by internal motion of partons within  the nucleon.

\section{Matrix Elements}

We  consider the double spin asymmetry $A_{LL}$ in  inclusive
charmonium state  production
 defined as
\begin{eqnarray}
A_{LL}^{H}(pp) = \frac{
d\sigma(p_+p_+\to H)-d\sigma(p_+p_-\to H)}
{d\sigma(p_+p_+\to H)+d\sigma(p_+p_-\to H}=
\frac{Ed\Delta\sigma/d^3p}{Ed\sigma/d^3p},
\end{eqnarray}
where $p_+(p_-)$ stands for the sign of the helicity projection  onto  
the proton momentum
direction and $H$ denotes the particular charmonium state.
The production of each quarkonium state receives  contributions from
both color singlet and color octet quark-antiquark pair states.
 We consider here only the dominant sets of
color octet states in the NRQCD velocity expansion for direct S and P
state charmonium production.
Hence we calculated the production asymmetries for the color octet
states 
 $^3P_{0,1,2}^8$, $^1S_0^8$ and $^3S_1^8$ together with those for
 color singlet states.

 The number of color octet long distance matrix elements can be reduced 
 using the NRQCD spin symmetry relations
\begin{eqnarray}
 \langle0|{\cal O}_8^H(^3P_J)|0\rangle &=& (2J+1)\langle0|{\cal O}_8^H(^3P_0)|0\rangle,\\
 \langle0|{\cal O}_8^{\chi_{cJ}}(^3S_1)|0\rangle &=& (2J+1)\langle0|{\cal O}_8^{\chi_{c0}}(^3S_1)|0\rangle,
\end{eqnarray}
which  are accurate up to $v^2$.

After having utilized these relations we remain with the three independent color
octet matrix elements
$\langle{\cal O}_8^{J/\psi}(^3S_1)\rangle$,
$\langle{\cal O}_8^{J/\psi}(^3P_0)\rangle$, and $\langle{\cal O}_8^{J/\psi}(^1S_0)\rangle$ which give the main
contributions to the direct $J/\psi$ hadroproduction cross
section. There is  only 
one color octet matrix element 
$\langle{\cal O}_8^{\chi_{c1}}(^3S_1)\rangle$
which contributes to $\chi_{cJ}$ states production at  lowest order.

As was mentioned above, the values of the matrix elements extracted from
the fixed target experiment's data contain large theoretical
uncertainties. In our further calculations we will use 
the following values for the 
three main matrix elements extracted from $J/\psi$ production at
CDF at the Tevatron using the GRV LO (1994) \cite{GRV}  parametrization for
unpolarized parton densities  \cite{BK}:
\begin{eqnarray}
 &&  \langle0|{\cal O}^{J/\psi}_8(^3S_1)|0\rangle= 1.06\pm0.14^{+1.05}_{-0.59}\cdot10^{-2}
    GeV^3,\\
&&\langle0|{\cal O}_8^{J/\psi}(^1S_0)|0\rangle+\frac{3.5}{m_c^2}\langle0|{\cal O}_8^{J/\psi}(^3P_0)|0\rangle =
3.9\pm1.15^{+1.46}_{-1.07}\cdot10^{-2} GeV^3
\end{eqnarray}
The second errors quoted in these expressions correspond
 to the variation of the factorization scale
$\mu$ from $0.5\sqrt{p_T^2+4m_c^2}$ to $2\sqrt{p_T^2+4 m_c^2}$.
From the errors indicated in (6) and (7) it is obvious that the variation
of the renormalization and/or factorization scale also leads to 
large uncertainties
when  fitting the color octet parameters. 

The fit results for  long distance color octet matrix elements
 can be affected also by
higher $v^2$ corrections.
In particular,  large uncertainties emerge when the matrix element
 $\langle O_8^{J/\psi}(^3S_1)\rangle$ is extracted from the data on
$J/\psi$ production at large $p_T$ at the Tevatron. In  fitting
  the CDF data the
energy of soft gluons emitted by a heavy quark-antiquark pair before
transition into $J/\psi$ is usually neglected \cite{CL,BK,CGMP} and,
consequently,  the
fragmentation function of a gluon into $J/\psi$ on the scale $2 m_c$ has
the form:
\begin{eqnarray}
D_{g\to\psi}(z,2m_c) = \frac{\pi\alpha_s(2 m_c)}{24 m_c^3}\delta(1-z)
\langle O_8^{J/\psi}(^3S_1)\rangle.
\end{eqnarray}
Here the delta function implies that the $J/\psi$ carries the whole energy
of the fragmenting gluon.
If the nonzero energy of the soft gluons
(of order $m_c v^2$) is taken into account the fragmentation
function becomes softer. Hence the realistic cross section 
is smaller at large $p_T$
and the fit value of the
$\langle O_8^{J/\psi}(^3S_1)\rangle$ matrix element appears about 
 a factor of
two larger \cite{MP,Bproc,ELV,BRW}.
Therefore,  care is required when using the value for this parameter
 extracted  from  CDF data at fixed target energies. Also, 
the uncertainty connected to the 'trigger bias' effect  makes it 
impossible to use this value for testing the NRQCD universality.

For  indirect $J/\psi$ production via decay of the underlying $\psi'$ state 
  and for production of $\chi_{cJ}$  states  
we use the following values fitted from  CDF data:
\begin{eqnarray}
\langle{\cal O}_8^{\chi_{c1}}(^3S_1)\rangle &= &9.8\cdot10^{-3} GeV^3 \cite{CL},\\
\langle{\cal O}_8^{\psi'}(^3S_1)\rangle &=& 0.46\cdot10^{-2} GeV^3 \cite{BK},\\
\langle{\cal O}_8^{\psi'}(^1S_0)\rangle+
\frac{3.5}{m_c^2}\langle{\cal O}_8^{\psi'}(^3P_0)\rangle &=&  1.8\cdot10^{-2} GeV^3
\cite{BK}.
\end{eqnarray}
For the calculation of the expected asymmetries we assume that the first
parameter is dominating in eq. (11). In this case the
discrepancy between the COM predictions and experimental data on  $\psi'$
polarization appears to be smallest \cite{Beneke}.

\section{Results and Discussion}
{\bf Asymmetries in production of quark-antiquark states.}
 The characteristic value of the partonic $x$ in the production of
$(c\bar c)$ pairs  can  be obtained from the relation
$x_1 x_2\simeq(4 m_c^2+p_T^2)/S$ ($\simeq 0.01$ at HERA-$\vec N$). 
This means that the typical values of $x_{gluon}$ which
 can  be probed by measuring the spin
 asymmetry in charmonium  production is about  $x_{gluon}\simeq0.1$. 
We used three  parametrizations for polarized parton distribution 
functions (PDF) that are different in the region $x\simeq0.1$
to show the dependence of spin asymmetries in the production of  various 
quark-antiquark pair states  on the gluon
polarization in the nucleon:
the old version of
the  Gehrmann-Stirling  (GS) parametrization
(set A) \cite{GSold} (as example of a large gluon polarization)
and the new version
of the  GS parametrizations in NLO and LO (both set A) \cite{GSnew}
(as examples for moderate gluon polarizations peaking at different
values of $x_{gluon}$). 
\begin{wrapfigure}{l}{7.5cm}
\vspace*{-12mm}
\centering
\epsfig{file=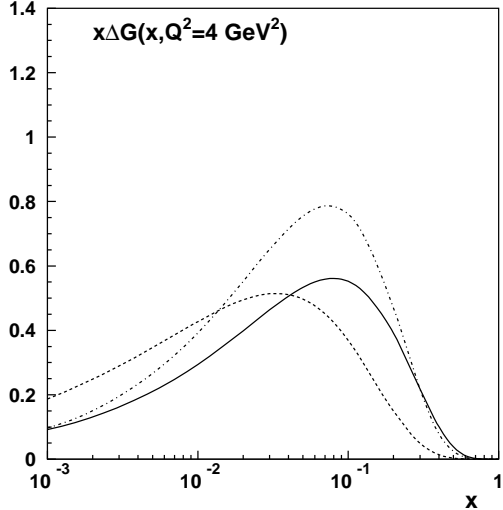,width=7.5cm}
\caption{\small Different possible polarized gluon
 distributions in the nucleon, used in this paper
 at $Q^2=4$ GeV:
 new GS parametrization NLO set A (solid line),
LO set A (dashed line) \cite{GSnew}  and  old
GS parametrization (set A) (dash-dotted) \cite{GSold}.
}
\end{wrapfigure}
 In  Fig.1 the polarized gluon densities from these
parametrizations are shown  at $Q^2=4$ GeV
(note that the polarized gluon 
distribution functions of NLO set B and  LO set A
practically coincide at this value of $Q^2$).
As can be seen from Fig.1, the three chosen sets exhibit  different 
values for the polarized gluon distribution function
for the partonic $x$ near $0.1$.
We note that, although the calculations of subprocess
level cross sections are performed in leading order, 
the  NLO set of the parametrization was
used   to
probe  different shapes of the polarized gluon distribution function.

  Fig.2  shows the expected
asymmetries for  different states of a heavy quark-antiquark
pair at $\sqrt{s}=40$ GeV (HERA-$\vec N$).
The asymmetries for $^1S_0^{(8)}$
 and $^3S_1^{(8)}$ octet states are represented by solid and dotted lines,
 respectively.
  The dashed lines correspond to  combined asymmetries of
 $^3P_J$ octet states, $\sum_{J=0,1,2}{(2J+1)^3P_J^{(8)}}$.
The dash-dotted lines show asymmetries of $J/\psi$ production in the CSM
including $J/\psi$ production through decays of higher charmonium states
($\psi'$ and $\chi_{cJ}$).
For unpolarized parton distribution functions  we used the GRV LO
parametrization \cite{GRV}.
\begin{figure}[ht]
\centering
\begin{minipage}[c]{4.9cm}
\centering
\epsfig{file= 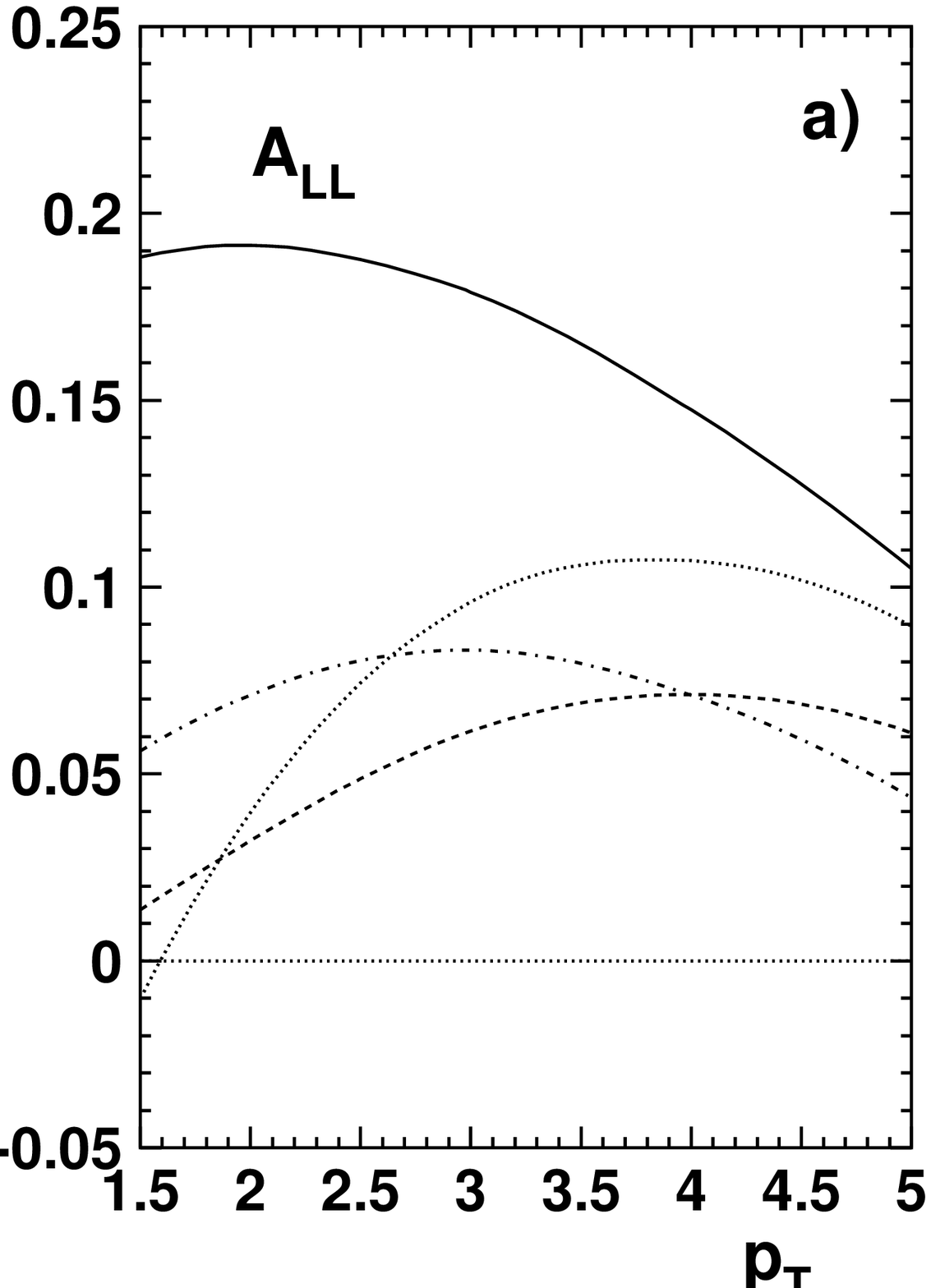,width=4.9cm}
\end{minipage}
\hspace*{0.4cm}
\begin{minipage}[c]{4.9cm}
\centering
\epsfig{file= 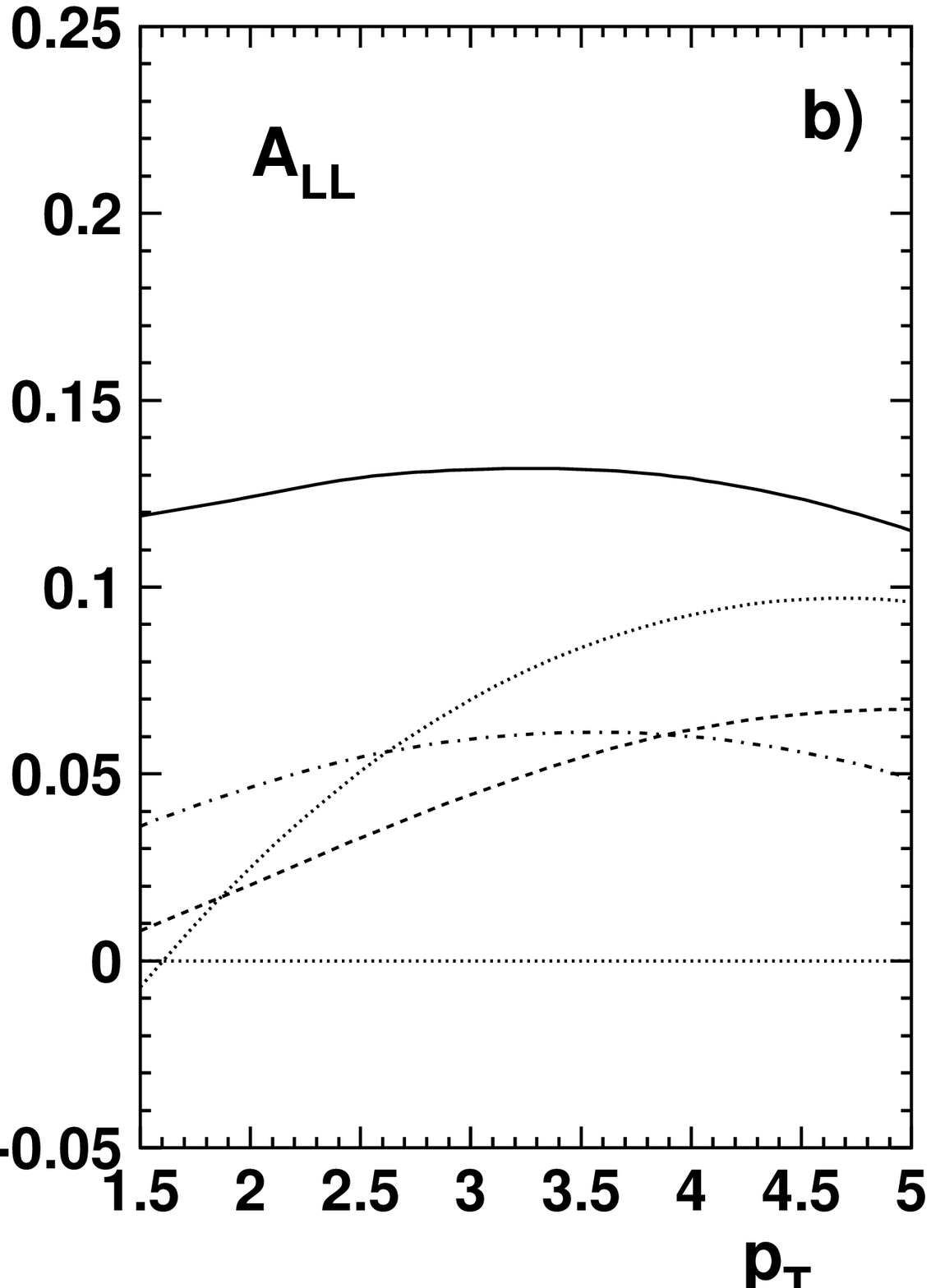,width=4.9cm}
\end{minipage}
\hspace*{0.4cm}
\begin{minipage}[c]{5cm}
\centering
\epsfig{file= 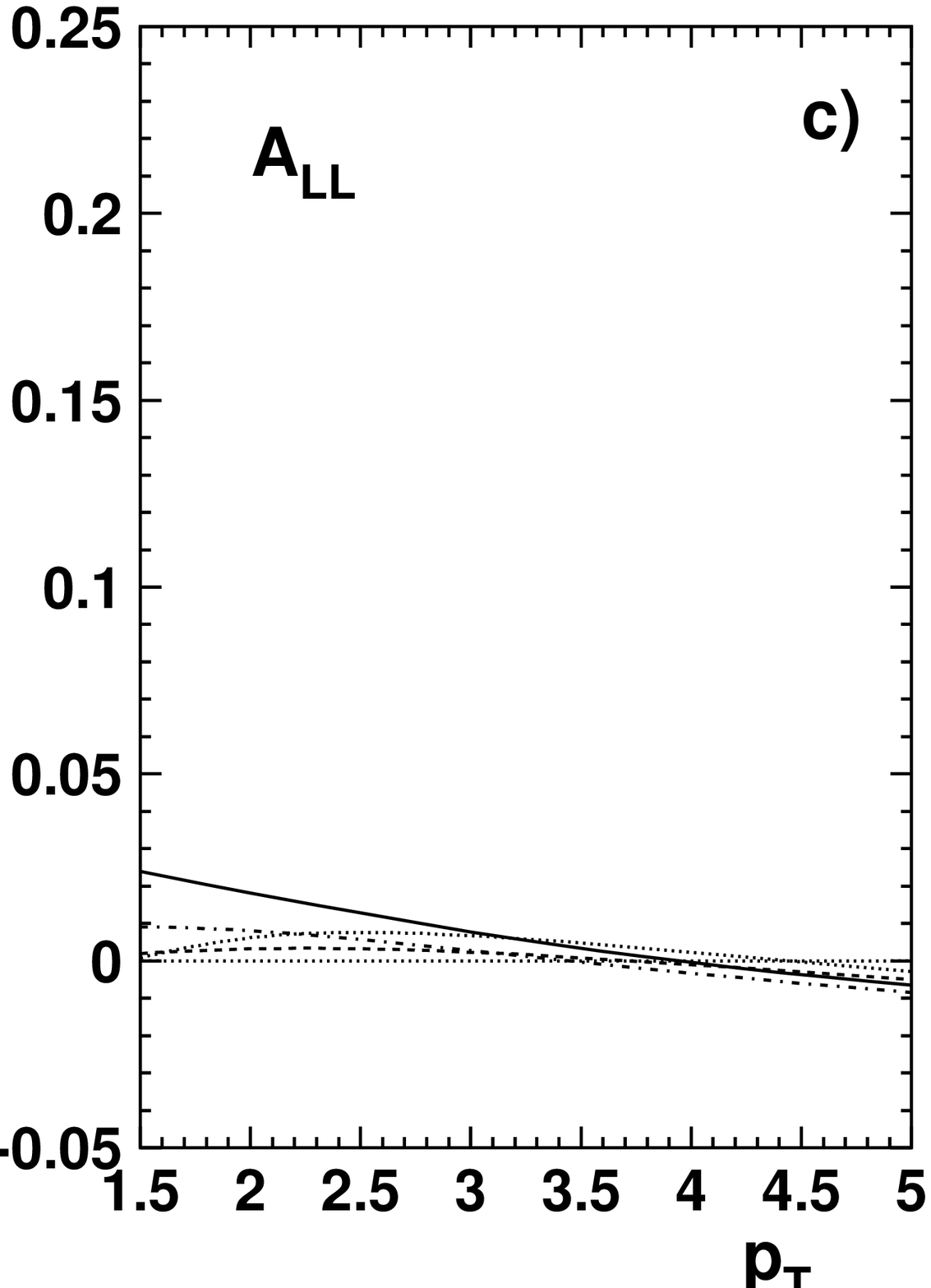,width=5.0cm}
\end{minipage}
\caption{\small
  The expected asymmetries at HERA-$\vec N$ ($\sqrt{s}=40$ GeV)
for the production of different color octet
states. Solid lines represent asymmetries for the 
$^1S_0^{(8)}$ state,
 dashed lines for the combination
of $^3P_J^{(8)}$ states (see text),
 dash-dotted lines for the $^3S_1^{(8)}$ octet state, and dotted
lines correspond to $J/\psi$ production in the CSM. The three figures
base upon different input distributions:
a) old GS parametrization 
(set A) \cite{GSold},  new version of
GS parametrization  b) NLO set A,  and c) LO set A
\cite{GSnew}.
}
\end{figure}

In all three parametrizations for the polarized PDF the
gluon-gluon fusion gives the
dominant contribution to  $\Delta\sigma$; 
quark-gluon subprocesses
contribute only to about $10\%$ and the contribution of
quark-antiquark annihilation subprocesses is less than $1\%$.

As can be seen from Fig.2, the expected asymmetries for all states
strongly depend on the size of the polarized gluon distribution function
in the region  $x_{gluon}\simeq0.1$.

\bf 
Measurement of the inclusive $J/\psi$ asymmetry. 
\normalsize 
In the inclusive case the kinematics of the $2\to2$ subprocess cannot
be reconstructed completely. Hence only indirect
information on the  gluon polarization in the nucleon
can be obtained by measuring the spin asymmetry in 
 inclusive $J/\psi$ production.

Figure 3 shows the expected double spin asymmetries at 
HERA-$\vec N$ energy
as a function of $J/\psi$ transverse momentum.
From now on we use the new GS parametrization, sets A,B, and C \cite{GSnew}.
For the mass of the charm quark the value $m_c=1.48$ GeV was taken and
the parton
distribution functions are evaluated on the factorization  scale
$\mu = \sqrt{p_T^2+4 m_c^2}$.  The strong coupling constant is calculated
by the one-loop formula with 4 active flavors ($\Lambda_{QCD}=200$ MeV).

Fig.3a corresponds to the case when the first parameter in 
the combination (7),
$\langle{\cal O}_8^{J/\psi}(^1S_0)\rangle$, is dominating,
i.e. $\langle{\cal O}_8^{J/\psi}(^3P_0)\rangle$=0.
 Fig.3b corresponds to the opposite case,
$\langle{\cal O}_8^{J/\psi}(^1S_0)\rangle$=0.
In  both cases we use for the third main
parameter the value
$\langle{\cal O}_8^{J/\psi}(^3S_1)\rangle=20\cdot10^{-3} GeV^3$
(the sensitivity of asymmetries to the value of this parameter will be
considered below).
\begin{figure}[ht]
\centering
\begin{minipage}[c]{7.5cm}
\centering
\epsfig{file= 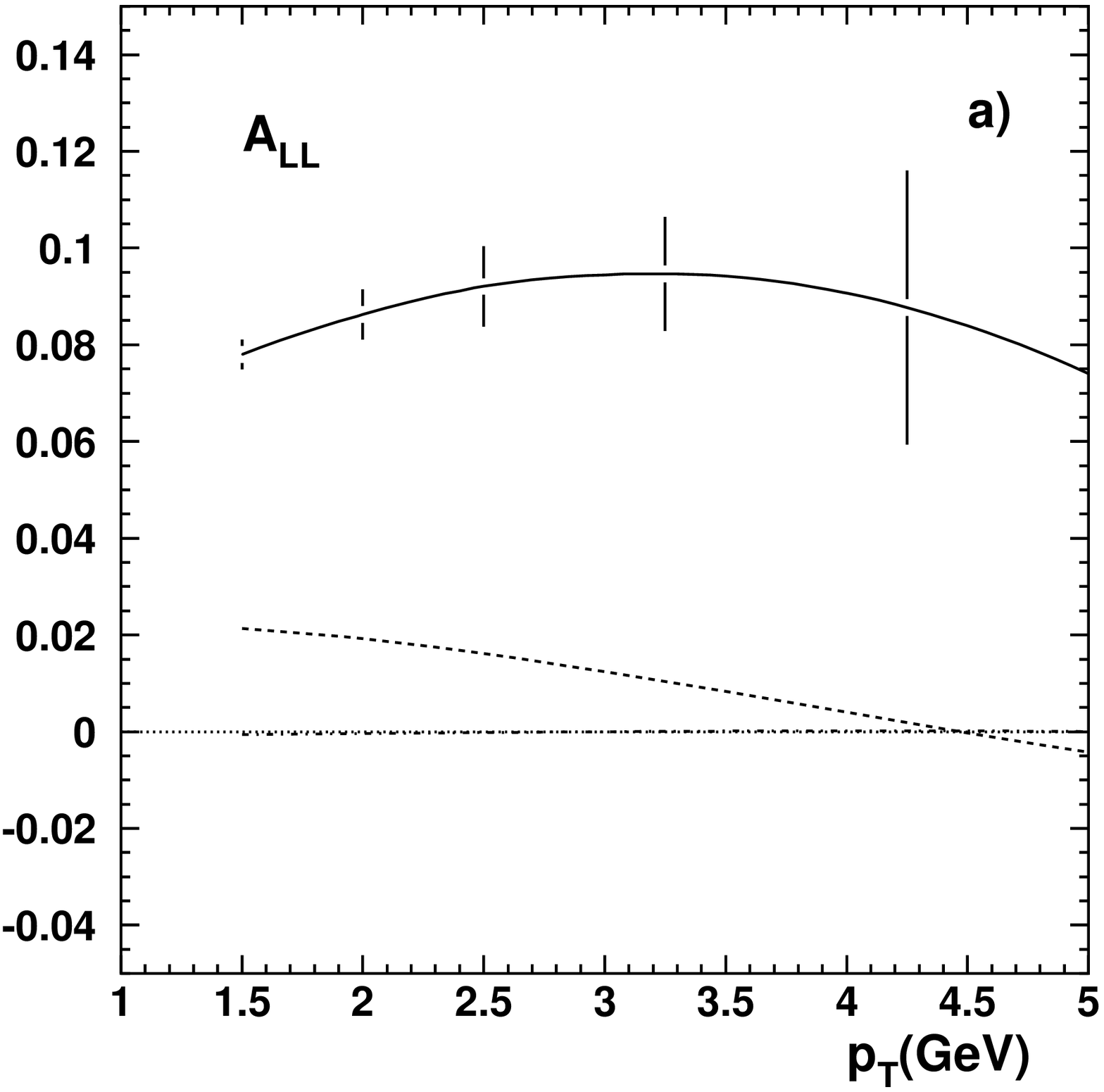,width=7.5cm}
\end{minipage}
\hspace*{0.5cm}
\begin{minipage}[c]{7.5cm}
\centering
\epsfig{file= 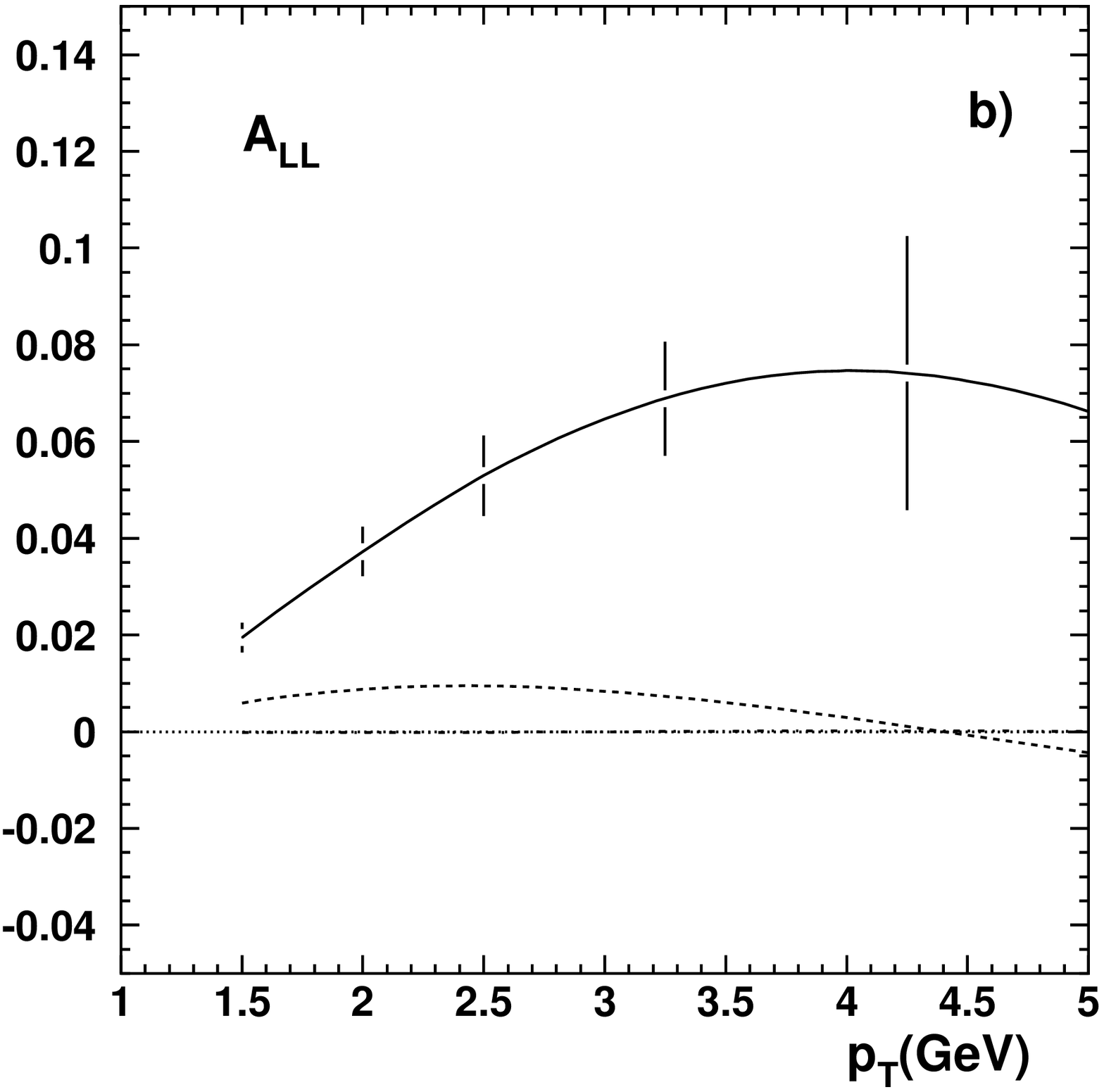,width=7.5cm}
\end{minipage}
\caption{\small
  The expected asymmetries versus transverse momentum 
at $\sqrt{s}=40$ GeV for  the NLO set A (solid lines), set B (dashed
lines) and set C (dash-dotted lines) of  the new
 GS parametrization \cite{GSnew};
a) $\langle{\cal O}^{J/\psi}_8(^3P_0)\rangle=0$, 
b) $\langle{\cal O}^{J/\psi}_8(^1S_0)\rangle=0$.
}
\end{figure}
Figures  3a and 3b also display the expected statistical errors.
The statistical error $\delta A_{LL}$ at HERA-$\vec N$
can be estimated from \cite{Nowak}
\begin{eqnarray}
\delta A_{LL} = 0.17/\sqrt{\sigma(pb)};
\end{eqnarray}
$100\%$ efficiency is assumed.
This relation has been determined by assuming an integrated luminosity of
$240~pb^{-1}$ and beam and target polarizations $P_B=0.6$ and 
$P_T=0.8$, respectively \cite{Nowak}.
The error bars  are obtained by using integrated cross sections over bins
$\Delta p_T=0.5$ GeV (for the first three points) and $\Delta p_T=1$ GeV (for
the other two ones).
The $J/\psi$ decay branching ratio into the $e^+e^-$ mode is also included.
As can be seen from Figs.3, the magnitude of
asymmetries and  expected errors allows one to distinguish
between different parametrizations of polarized parton distribution
functions.

\bf Production asymmetry of  $\chi_{c1}$ and $\chi_{c2}$ states. 
\normalsize
In  Figs.4a,b we show the production asymmetries of 
$\chi_1$ and $\chi_2$ states 
in conjunction with the projected HERA-$\vec N$ errors for the
 three different sets  used for polarized PDF.
Unlike direct  $J/\psi$ production the production asymmetries of 
$\chi$ states  depend    only on one color octet matrix
element, namely $\langle{\cal O}_8^{\chi_{c1}}(^3S_1)\rangle$. 
The value of this parameter was extracted from Tevatron data at large
$p_T$ values, i.e. from the region where the NRQCD factorization
mechanism is valid, and higher twist effects are expected to be small, 
i.e. of the order of $\Lambda/p_T$ or even $(\Lambda/p_T)^2$.
Hence  the production asymmetries of  $\chi$ states  do not contain 
additional free parameters and enable us to probe
the polarized gluon distribution function
with less theoretical uncertainties. 
We note that  the branching ratios of $\chi_J$  decay 
into $J/\psi$ plus photon are taken into account in  the
calculations of the expected errors.  

\begin{figure}[ht]
\centering
\begin{minipage}[c]{7.5cm}
\centering
\epsfig{file= 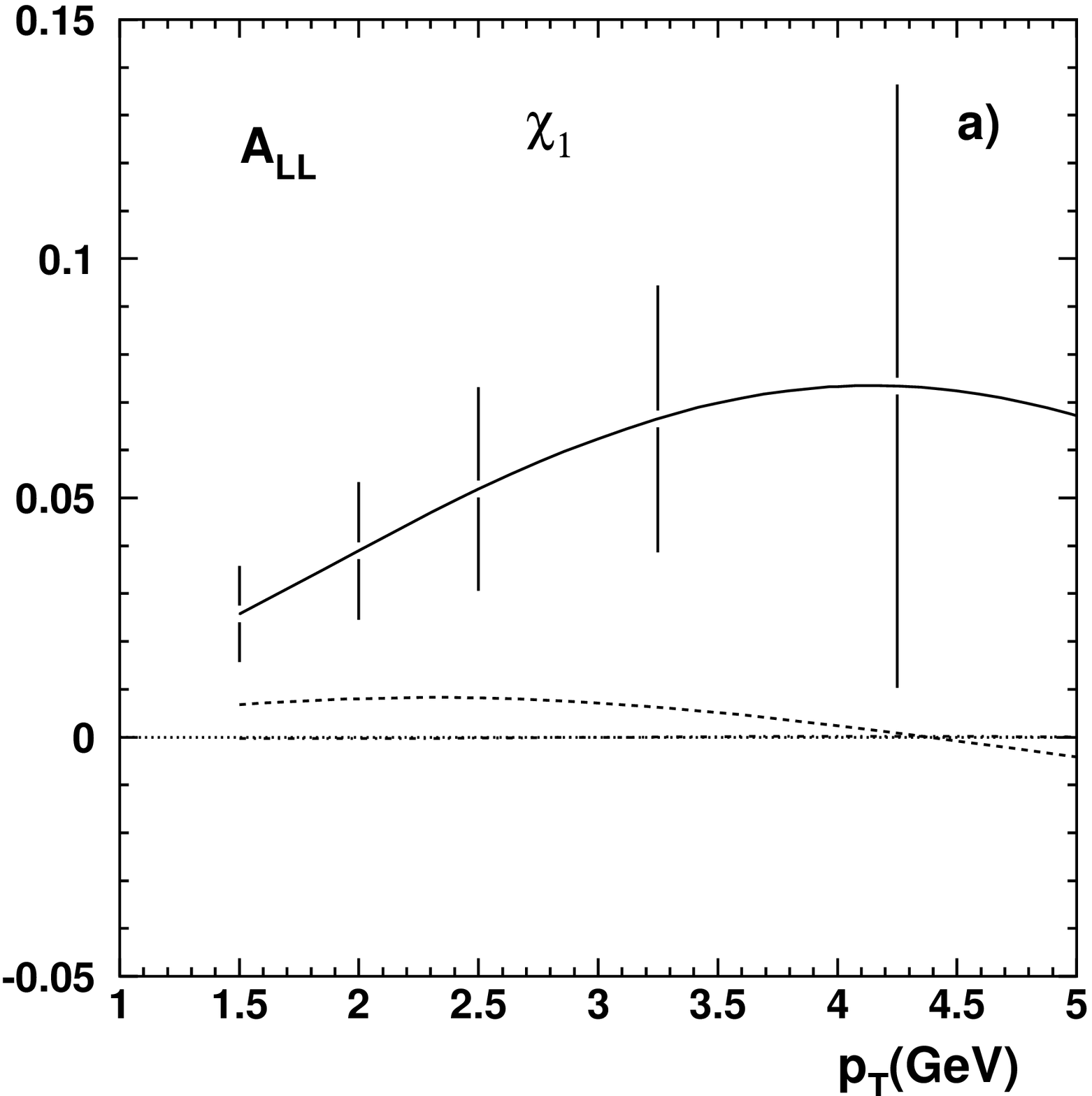,width=7.5cm}
\end{minipage}
\hspace*{0.5cm}
\begin{minipage}[c]{7.5cm}
\centering
\epsfig{file= 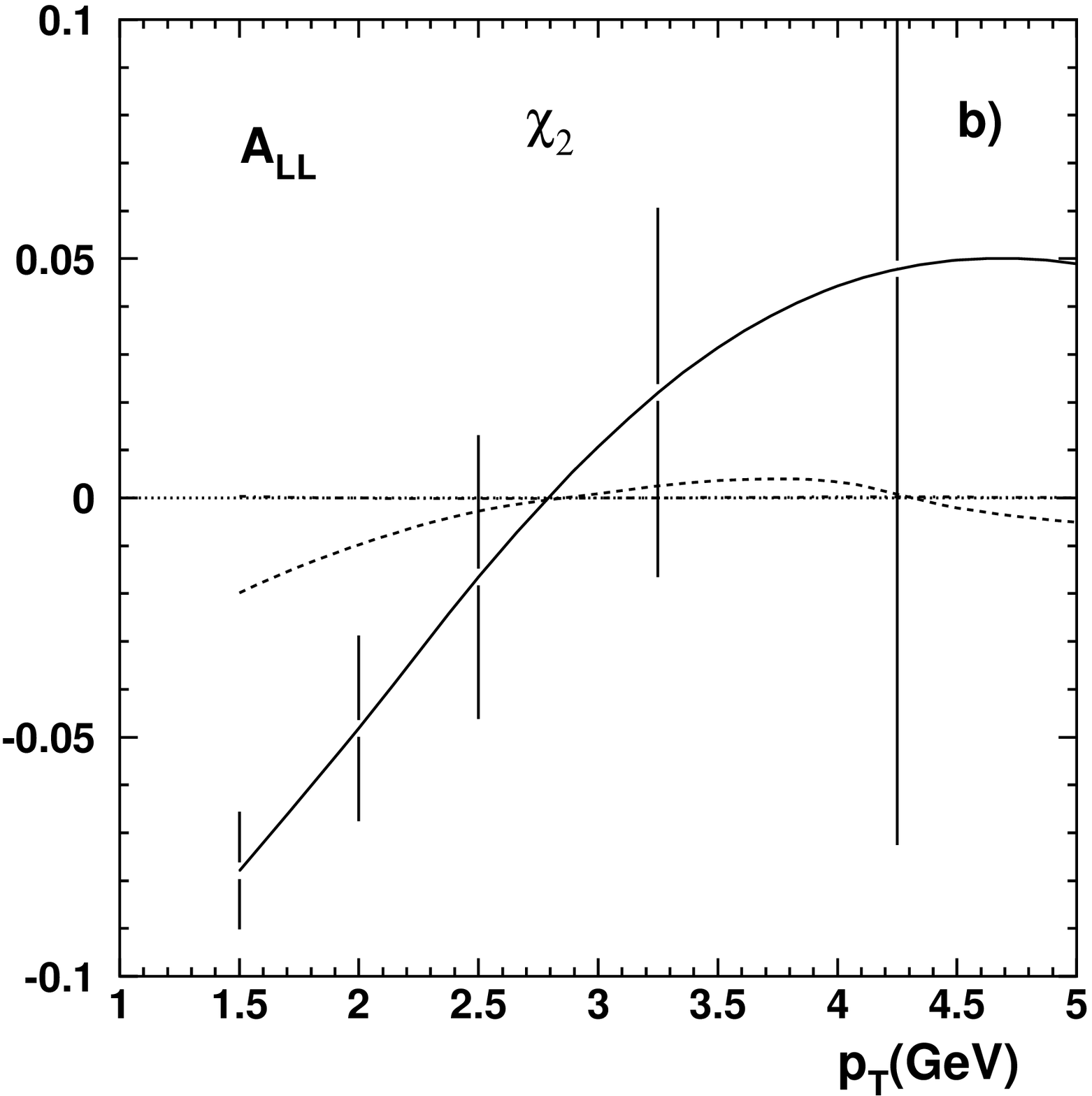,width=7.5cm}
\end{minipage}
\caption{\small
  The $\chi_{cJ}$ production  asymmetries versus transverse momentum 
at $\sqrt{s}=40$ GeV for  the NLO set A (solid lines), set B (dashed
lines) and set C (dash-dotted lines) of  the new
 GS  parametrization \cite{GSnew};
 a)  $\chi_{c1}$, b)$\chi_{c2}$.
}
\end{figure}

\bf 
$J/\psi$  plus jet production.
\normalsize
 Direct access to the
ratio $\Delta G(x)/G(x)$ is possible only if the 'other jet' 
is detected  ({\it photon+jet} or $J/\psi$+{\it jet}) \cite{Ans},
because the complete
kinematics of the $2\to2$ subprocess can be reconstructed if the away-side
jet in the production of $J/\psi$ is measured, as well.
If the values of the long distance matrix elements are established from
experiments where the uncertainties connected with higher twist
and other corrections are expected
 to be small (or negligible), the $J/\psi$+{\it jet}
production asymmetry at HERA-$\vec N$ will serve as a good
 tool for the extraction of the
$\Delta G(x)/G(x)$ value at $x\simeq0.1$  \cite{Ans}.

{\bf  Measurement of color octet matrix elements}. 
As soon as  the polarized gluon distribution function is  extracted
from other channels  (e.g. {\it photon+jet})
 at HERA-$\vec N$ \cite{Ans} or RHIC \cite{RHIC},
a measurement of the  $J/\psi$ asymmetry can be used for
checking the  NRQCD factorization scheme.

For further calculations and analysis we choose
\begin{figure}[ht]
\centering
\begin{minipage}[c]{7.5cm}
\centering
\epsfig{file= 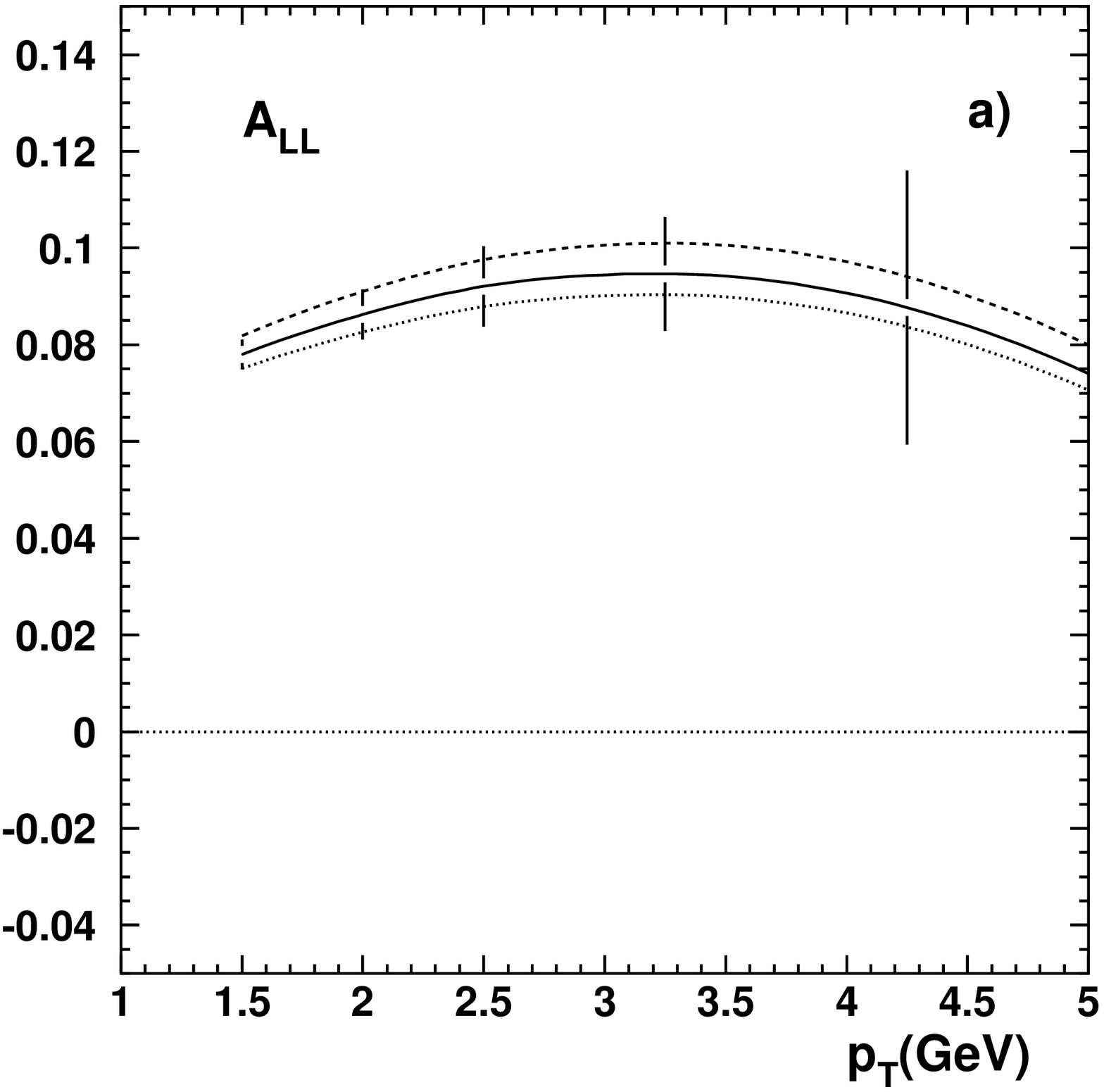,width=7.5cm}
\end{minipage}
\hspace*{0.5cm}
\begin{minipage}[c]{7.5cm}
\centering
\epsfig{file= 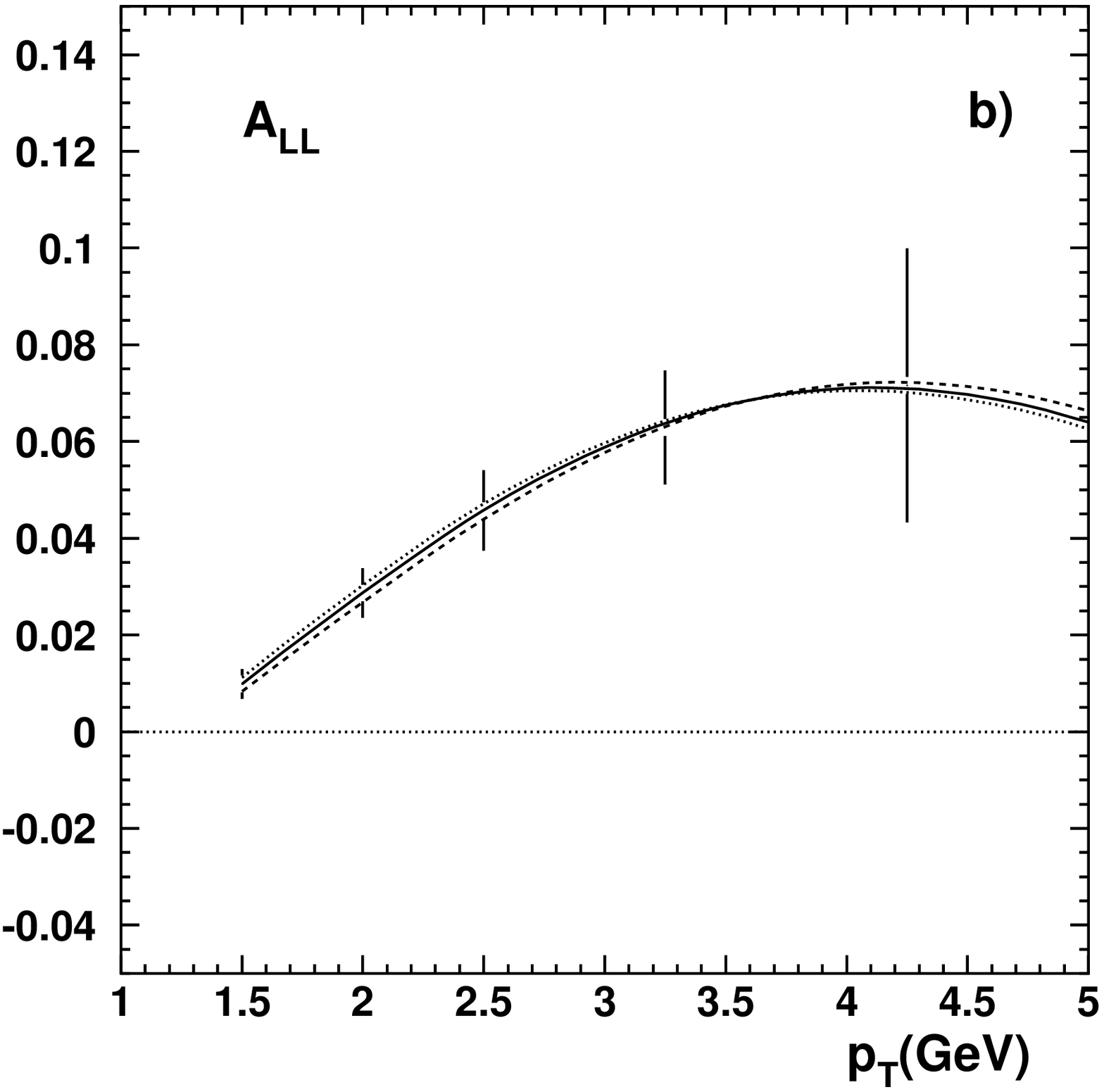,width=7.5cm}
\end{minipage}
\caption{\small
 The expected asymmetries at HERA-$\vec N$ energy for different values 
of  the long-distance parameter $\langle{\cal O}_8^{J/\psi}(^3S_1)\rangle$. 
The solid lines correspond to the value 
$20\cdot10^{-3}$ $GeV^3$, the 
dashed lines  to $10\cdot10^{-3}$ $GeV^3$,  and the dotted lines 
to $30\cdot10^{-3}$ $GeV^3$;
a)  asymmetries for the case
when 
 $\langle{\cal O}^{J/\psi}_8(^3P_0)\rangle=0$, 
b) for $\langle{\cal O}^{J/\psi}_8(^1S_0)\rangle=0$.
}
\end{figure}
 the new  GS
parametrization for polarized PDF (NLO, set A) \cite{GSnew}.
In figures 5a and 5b we  present the expected asymmetries in  $J/\psi$
production for two radical choices of
$\langle{\cal O}_8^{J/\psi}(^1S_0)\rangle$ and
$\langle{\cal O}_8^{J/\psi}(^3P_0)\rangle$.
Figure 5a corresponds to the case
  when the parameter
$\langle{\cal O}^{J/\psi}_8(^3P_0)\rangle$ tends to zero, 
i.e. $\langle{\cal O}^{J/\psi}_8(^1S_0)\rangle$ is the
dominating term in the combination (7). Figure 5b
 represents the other radical choice, when the  second parameter
$\langle{\cal O}^{J/\psi}_8(^1S_0)\rangle$ is zero.
As already mentioned above, due to the 'trigger bias' effect 
the value of the third main parameter,
$\langle{\cal O}_8^{J/\psi}(^3S_1)\rangle$,   is
underestimated about twice. To check the sensitivity
of
\begin{wrapfigure}{l}{7.5cm}
\vspace*{-12mm}
\centering
\epsfig{file= 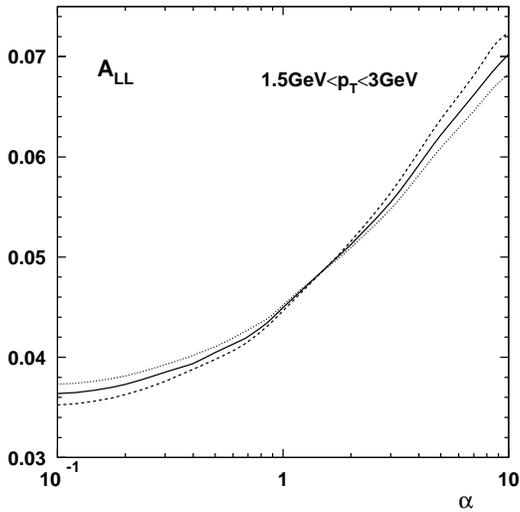,width=7.5cm}
\caption{\small
 The integrated double spin asymmetry
versus the ratio of color octet parameters,
$\alpha = \langle{\cal O}_8^{J/\psi}(^1S_0)\rangle m_c^2/
\langle{\cal O}_8^{J/\psi}(^3P_0)\rangle$. Solid line: 
$\langle{\cal O}_8^{J/\psi}(^3S_1)\rangle=20\cdot10^{-3}$ $GeV^3$,
dashed line: $10\cdot10^{-3}$ $GeV^3$,  dotted line: 
$30\cdot10^{-3}$ $GeV^3$. 
}
\end{wrapfigure}
 the expected asymmetries to the uncertainty caused by this effect,
we used three different values for
$\langle{\cal O}_8^{J/\psi}(^3S_1)\rangle$,
$10\cdot10^{-3}$ GeV$^3$ from (6) (dashed lines in  Figs.5),
 $20\cdot10^{-3}$ GeV$^3$ (solid lines) and $30\cdot10^{-3}$ GeV$^3$
 (dash-dotted lines).
From comparison of Figs.5a and b it can be seen that
   the asymmetries depend strongly on the
  relative values of the matrix elements
$\langle{\cal O}_8^{J/\psi}(^1S_0)\rangle$ and
$\langle{\cal O}_8^{J/\psi}(^3P_0)\rangle$
and practically do not depend on the value of
$\langle{\cal O}_8^{J/\psi}(^3S_1)\rangle$.
Hence, independently of the value of the latter parameter and, consequently,
independently of the   'trigger bias' effect,
a  measurement of the $J/\psi$
production asymmetry can be used for an 
extraction of  the ratio of the first two
parameters.

In  Fig.6 we plot the expected integrated asymmetry ($1.5<p_T<3$ GeV)
versus  the ratio
$\alpha =m_c^2 \langle{\cal O}_8^{J/\psi}(^1S_0)\rangle/
\langle{\cal O}_8^{J/\psi}(^3P_0)\rangle$
at  HERA-$\vec N$ energy
for the  above presented three different values of the third parameter.
The expected statistical error for the integrated asymmetry is about
0.002, i.e. it will be possible to determine the ratio $\alpha$ within
about $\pm15\%$ if $\alpha$ is larger than unity. With $\alpha$ being
smaller the sensitivity is becoming worse.

 Of course, this method requires that 
the polarized
gluon distribution function is known already and  large
 enough to generate   observable
sizes of $J/\psi$ production asymmetries.

One of the main parameters of the factorization approach is the mass of the charm quark.
As was shown in \cite{TT}, the expected asymmetries are
practically insensitive to the value of the  charm quark mass.
Therefore, the double spin asymmetry in  $J/\psi$ production,
unlike the cross section, should be free from uncertainties caused
by the unknown mass of  intermediate color octet states.
We also note that the asymmetries do not depend strongly on the renormalization
scale unlike the $J/\psi$ production cross section \cite{BK}.

Finally, we note that recently 
 a possibility was shown to extract the 
  ratio $\alpha$ 
of color octet long distance parameters from
polarized $J/\psi$ electroproduction data with 
reasonable errors \cite{FleN}. 

\section{Conclusions}
We investigated the expected double spin asymmetries in
heavy quarkonium hadroproduction in polarized proton proton
collisions.
To reduce the contribution from possible higher twist
 corrections \cite{VHBT} we
considered     $J/\psi$ meson production at nonzero transverse momenta,
$p_T>1.5$ GeV. Unlike the calculations of \cite{GM}, where only the lowest
order subprocesses, $2\to1$, were taken into account, we
considered $J/\psi$ production in   $2\to2$ subprocesses at large
enough $p_T$ to avoid
 uncertainties coming from higher twist corrections \cite{VHBT}.

 The size of the expected asymmetries in conjunction with the 
 statistical errors at HERA-$\vec N$ allow  to distinguish
between different parametrizations for polarized PDF
 (GS, set A, B and C \cite{GSnew}). If, on
the  other hand, those were already known
measuring the asymmetry  would give a possibility to extract information
about the color octet long distance matrix elements
$\langle{\cal O}_8^{J/\psi}(^1S_0)\rangle$ and
$\langle{\cal O}_8^{J/\psi}(^3P_0)\rangle$,
separately. This would deliver useful information 
to check the universality of the factorization scheme.

A.T. acknowledges  the partly support of this work by the Alexander von
Humboldt Foundation. O.T. and A.T.  wish to thank the Organizers of 
the Workshop and
all the participants for the warmful atmosphere felt during their
stay in Zeuthen.

\end{document}